\documentclass[aps,pra,twocolumn]{revtex4}

\usepackage{graphicx}
\usepackage{pstricks}
\usepackage{bm}
\usepackage{amsfonts}
\newcommand{\mb}{\mathbf}
\newcommand{\be}{\begin{equation}}
\newcommand{\ee}{\end{equation}}
\newcommand{\bwt}{\begin{widetext}}
\newcommand{\ewt}{\end{widetext}}

\begin{document}

\title{Quantum states for Heisenberg limited interferometry}
\author{H.~Uys and P.~Meystre}
\affiliation{Department of Physics and College of Optical Sciences\\The University of Arizona, Tucson, AZ, 85721}
\begin{abstract}

The phase sensitivity of interferometers is limited by the
so-called Heisenberg limit, which states that the optimum phase
sensitivity is inversely proportional to the number of interfering
particles $N$, a $1/\sqrt{N}$ improvement over the standard
quantum limit. We have used simulated annealing, a global
optimization strategy, to systematically search for quantum
interferometer input states that approach the Heisenberg limited
uncertainty in estimates of the interferometer phase shift.  We
compare the performance of these states to that of other
non-classical states already known to yield Heisenberg limited
uncertainty.

\end{abstract}
\maketitle

\section{Introduction}

An important aspect of quantum metrology is the engineering of
quantum states with which to achieve measurements whose precision
is Heisenberg limited. In this limit the measurement
uncertainty is inversely proportional to the number of interfering
particles $N$, representing a $1/\sqrt{N}$ improvement over the
standard quantum limit. Squeezed light has long been employed to
beat the shot-noise limit \cite{Grangier1987,Xiao1987} and a
growing body of theoretical literature indicates that the
Heisenberg limit is in principle achievable using more exotic
quantum states as interferometer inputs
\cite{Yurke1986a,Holland1993,Sanders1995,Kim1998,Combes2005,Pezze2006}.
Several proof-of-principle experimental realizations of such
states have recently been carried out
\cite{Molmer1999,Sackett2000,Leibfried2004,Walther2004,Mitchell2004}. Other proposals to
beat the standard quantum limit involve the use of feedback
schemes \cite{Berry2001,Denot2006} or multi-mode interferometry
\cite{Soderholm2003}.  The potential superiority of atomic
fermions over bosons in some applications of atom interferometry
with quantum-degenerate atomic gases has also been pointed out
\cite{Search2003,Javanainen2007}.

This paper summarizes the results of a systematic search for input
quantum states that lead to Heisenberg limited interferometric
detection of phase shifts. Using the global optimization method of
simulated annealing we demonstrate the existence of numerous
possibilities over-and-above those already proposed in the
literature, and we evaluate and compare their performance.

Section II discusses our theoretical model of a Mach-Zehnder
interferometer used to measure the relative phase shift $\phi$
accumulated during the propagation of single-mode optical or
matter waves along its two arms. Section III introduces a
likelihood function used to estimate that phase and discusses its
asymptotic form in the limit of many measurements. Section IV
summarizes our main results obtained using simulated annealing and
section V focuses on the prospects for the experimental
realization of a quantum state of particular interest. Finally,
section VI is a summary and conclusion.

\section{Mach-Zehnder Interferometer}

\begin{figure}
\includegraphics[angle=0, scale=0.6]{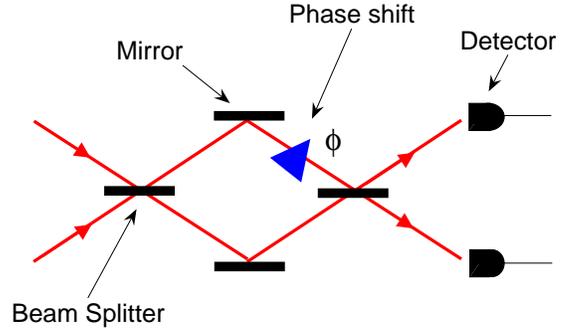}
\caption{(Color online) Schematic of a Mach-Zehnder interferometer, with
relative phase shift $\phi$ resulting from propagation of an input
field through its arms.}\label{mz}
\end{figure}

We consider a Mach-Zehnder interferometer with two input ports $A$
and $B$, see Fig.~\ref{mz}, characterized by bosonic annihilation
and creation operators $\hat a$ and $\hat a^\dagger$ and $\hat b$ and
$\hat b^\dagger$, respectively. We restrict our investigation to a system with fixed particle number $N$, in which case its
properties are conveniently described in terms of the angular momentum operators
 \cite{Sanders1995}
\begin{eqnarray}
\label{jx}
\hat{J}_x &=& \frac{\hat a^\dagger \hat b + \hat b^\dagger \hat a}{2},\\
\hat{J}_y &=& \frac{\hat a^\dagger \hat b - \hat b^\dagger \hat a}{2i},\\
\hat{J}_z &=& \frac{\hat a^\dagger \hat a - \hat b^\dagger \hat b}{2},\\
\hat{J}^2 &=& \hat{J}_x^2+\hat{J}_y^2+\hat{J}_z^2,
\end{eqnarray}
which obey the familiar commutation relations $\left[\hat J_i,\hat
J_j\right]=\varepsilon_{ijk}i\hat J_k$, where $\varepsilon_{ijk}$
is the Levi-Civita symbol, and $\left[\hat J^2,\hat J_i\right]=0$.
Choosing $z$ as the quantization axis we work in the basis of
eigenstates common to $\hat J^2$ and $\hat J_z$,
    \be
    |j,m\rangle_z\equiv|n_a\rangle|n_b\rangle.
    \ee
Here $|n_i\rangle$ is a Fock state with $n_i$ particles in arm
$I=A,B$. For brevity we drop the subscript $z$ henceforth. The
eigenvalues corresponding to $\hat J^2$ and $\hat J_z$ are
$j(j+1)$ and $m$ respectively, where
    \be
    j=(n_a+n_b)/2=N/2
    \ee
and
    \be m=(n_a-n_b)/2.
    \ee
In terms of these operators, the propagation of the input fields
through the interferometer -- which consists of three unitary
transformations describing an input $50/50$ beam-splitter, the
relative phase shift $\phi$, and an output $50/50$ beam splitter
-- reads \cite{Sanders1995}
    \begin{eqnarray}
    |\psi_{\rm out}\rangle &=& e^{-i(\pi/2)\hat J_x}
    e^{i\phi\hat J_z}e^{i(\pi/2)\hat J_x}|\psi_{\rm in}\rangle\\
    &=& e^{-i\phi\hat J_y}|\psi_{\rm in}\rangle.
\end{eqnarray}
Our aim is to find input states
    \be \label{instate}
    |\psi_{\rm in}\rangle =  \sum_{m=-N/2}^{N/2} \alpha_m|j,m\rangle
    \ee
such that the uncertainty in the estimate of the phase $\phi$ is
minimized.

Our restriction in this paper to fixed \textit{total} particle number leads to considerable analytical and computational simplification, but the more general problem in which the \textit{average} particle number is conserved is also of interest, and can in principle be carried out with the same techniques.

\section{Likelihood Function}

A number of approaches have been used as measures of the uncertainty in estimate of the relative phase $\phi$.
Commonly the standard error propagation formula is used to express this phase uncertainty in terms of the mean square error of a measured observable such as the particle number difference \cite{Search2003}
\be
\Delta\phi = \Delta\hat J_z/(d\hat J_z/d\phi).
\label{standerr}
\ee
Probability operator measures are also used \cite{Berry2001}, as well as information theoretical measures such as the Shannon mutual information \cite{Bahder2006}. 
In this paper we estimate the relative phase following an
operational approach based on Bayes' theorem
\cite{Braunstein1992,Holland1993,Hradil1995,Hradil1996}.  Consider
an experiment in which the probability amplitude of the $i^{\rm
th}$ basis state of an $N+1$ dimensional Hilbert space depends on
some phase $\phi$,
\begin{equation}
|\psi\rangle = \sum_{i=0}^N \alpha_i(\phi)|i\rangle.
\end{equation}
The probability to measure $|i\rangle$ conditioned on that phase
is $P(i|\phi) = |\alpha_i(\phi)|^2\label{probofi}$,
with
\be
\sum_{i=0}^N P(i|\phi) = 1.
\label{nrmPiphi}
\ee

Bayes' theorem states that the probability that the phase shift
has the value $\phi$, conditioned on the outcome $i$, is
\begin{equation}
\label{bayes}
P(\phi |i) = \frac{P(\phi)P(i|\phi)}{P(i)},
\end{equation}
where $P(\phi)$ is the phase probability distribution
\textit{prior} to the measurement and $P(i)$ is the prior
detection probability for the outcome $i$. Following a measurement
with outcome $i_1$, the phase probability distribution becomes
$P(\phi|i_1)$, which may now be used as the prior phase
probability distribution for a second measurement
\cite{Holland1993}, so that
    \begin{equation}
    \label{bayes2}
    P(\phi |i_1,i_2) = \frac{P(\phi|i_1)P(i_2|\phi)}{P(i_2)}
    = \frac{P(\phi)P(i_1|\phi)P(i_2|\phi)}{P(i_1)P(i_2)}.
    \end{equation}
Likewise, the phase probability distribution conditioned on the
outcome of a sequence of $M$ measurements is
    \begin{equation}
    \label{bayesiter}
    P(\phi |i_1,i_2,...,i_M) = \frac{P(\phi)P(i_1|\phi)
    P(i_2|\phi)...P(i_M|\phi)}{P(i_1)P(i_2)...P(i_M)}.
\end{equation}
For a large number of measurements, $M \gg 1$, and assuming that
the true phase shift is $\phi=\theta$, the number of times a
factor $P(i|\phi)$ appears in the product (\ref{bayesiter}) is
approximately $P(i|\theta)M$. This motivates the introduction of a
likelihood function for the phase shift to be $\phi$, conditioned
on its true value being $\theta$, as \cite{Hradil1995,Hradil1996}
    \be \label{LF}
    P_M(\phi |\theta) = \frac{1}{\cal{N}}\prod_{i=0}^N
    P(i|\phi)^{P(i|\theta) M},
    \ee
where
    \begin{equation}
    \mathcal{N} = \int_{-\frac{\pi}{2}}^{\frac{\pi}{2}} d\phi^\prime\prod_{i=0}^N
    P(i|\phi^\prime)^{P(i|\theta)M}
    \label{nrm}
    \end{equation}
is a normalization constant.  The likelihood function $P_M(\phi
|\theta)$ has the desirable property that it possesses a maximum
at the true value, $\theta$, of the phase shift.  This is easily
shown by taking its derivative
\begin{widetext}
\begin{eqnarray}
\nonumber
\frac{dP_M(\phi|\theta)}{d\phi} &=& \sum\limits_{i=0}^N
\left[MP(i|\theta)P(i|\phi)^{\left[
MP(i|\theta)-1\right]}\frac{dP(i|\phi)}{d\phi}
\prod_{k\neq i}^NP(k|\phi)^{MP(k|\theta)}\right]\\
&=&M\left[\prod_{k=0}^NP(k|\phi)^{MP(k|\theta)}\right]\sum\limits_{i=0}^N
\frac{P(i|\theta)}{P(i|\phi)}\frac{dP(i|\phi)}{d\phi}.
\label{dpdphi}
\end{eqnarray}
\end{widetext}
Evaluating Eq. (\ref{dpdphi}) at $\theta$, together with the
normalization condition (\ref{nrmPiphi}) gives then
\begin{equation}
\frac{dP(\phi|\theta)}{d\phi}\left.\vphantom{\frac{}{}}\right|_\theta =
M\left[\prod_{k=0}^NP(k|\theta)^{MP(k|\theta)}\right]\sum\limits_{i=0}^N
\frac{dP(i|\phi)}{d\phi}\left.\vphantom{\frac{}{}}\right|_\theta=0,
\end{equation}
implying an extremum at $\theta$. Taking the second derivative and
again using normalization shows this extremum to be a maximum.

In order to estimate the phase uncertainty in the limit of large
$M$ we introduce the function
    \be
    K(\phi|\theta) = \log{P(\phi|\theta)}.
    \ee
Expanding then $K(\phi|\theta)$ around $\phi=\theta$ and
accounting for the normalization condition (\ref{nrmPiphi}) we find
that $P(\phi|\theta)$ is approximately given by
    \be
    P(\phi|\theta)\approx
    e^{K(\theta|\theta)-\frac{(\phi-\theta)^2}{2\sigma^2}}
    \label{expandedent}
    \ee
where
\be
    \sigma^2 = \frac{1}{M\sum_{i=0}^N\left(\frac{d
    P(i|\phi)}{d\phi}\vert_\theta\right)^2/P(i|\theta)}=\frac{1}{MF},
    \label{fisher}
    \ee
and
    \be
    F=\sum_{i=0}^N\left(\frac{d P(i|\phi)}{d\phi}
    \Big\vert_\theta\right)^2/P(i|\theta) \label{deffisher}
    \ee
is the so-called Fisher information, as shown in Appendix A
\cite{Cover2006}. For large $M$ the exponential suppresses
strongly those contributions to $P(\phi|\theta)$ for which $\phi
\gtrsim \sigma$ so that $P(\phi|\theta)$ becomes Gaussian in that
limit \cite{Braunstein1992}.  Asymptotically, the likelihood
function is therefore completely characterized by its variance, or
equivalently by its Fisher information.  Equation (\ref{fisher})
also shows that the phase uncertainty decreases as the inverse
square root of the number of measurements.

The Fisher information plays an important role in information
theory as it gives a lower limit to the variance of any estimator
via the Cramer-Rao inequality \cite{Cover2006}
    \be
    {\rm var}(x)\ge\frac{1}{F}, \label{cramerrao}
    \ee
where ${\rm var}(x)$ is the mean square error of the random
variable $x$ being estimated and Eq. (\ref{cramerrao}) is the defining relation for the Fisher information. (Note also that the Fisher
information of $M^\prime$ independent and identically distributed
samples is $M^\prime$ times the individual Fisher information.)
Thus Eq.~(\ref{expandedent}) indicates that the likelihood
function $P_M(\phi|\theta)$ achieves the Cramer-Rao limit. It
permits us to find input states of the interferometer of the form
of Eq.~(\ref{instate}) that result in an estimate of the phase
shift with minimum uncertainty.

To illustrate how the likelihood function may be used to estimate
a phase shift experimentally, consider a thought experiment using
the Mach-Zehnder interferometer in Fig. 1.  Each measurement
counts the number of particles $n_a$ exiting the interferometer in
arm ``A''. Due to particle conservation, this is a direct measure
of the quantum number $m$. Expanding the exit state of the field
as
    \begin{equation}
    |\psi_{\rm out}(\phi)\rangle =
    \sum_m\alpha_m(\phi)|j,m\rangle,
    \end{equation}
each measurement yields a specific particle number $n^{(i)}_a=
N/2+ m^{(i)}$ with an associated phase probability distribution
    \begin{equation}
    P(m^{(i)}|\phi) = |\alpha_{m}^{(i)}(\phi)|^2.
    \end{equation}
After $M$ such measurements the conditional phase probability
distribution takes the form of Eq. (\ref{bayesiter}), which for
sufficiently large $M$ is a good approximation to the likelihood
function Eq. (\ref{LF}) --- up to the normalization constant as in
Eq. (\ref{nrm}).  The maximum of this conditional phase
probability distribution is an estimate of the phase shift and its
variance gives the uncertainty.

An important consideration is the number of measurements needed
for the conditional phase probability distribution, Eq.
(\ref{bayesiter}), to be an accurate representation of the
likelihood function.  This matter is not addressed in this paper where we use throughout the assymptotic form of the likelihood function, but has been investigated by Braunstein \cite{Braunstein1992b}.

\begin{figure*}
\includegraphics[angle=0, scale=0.6]{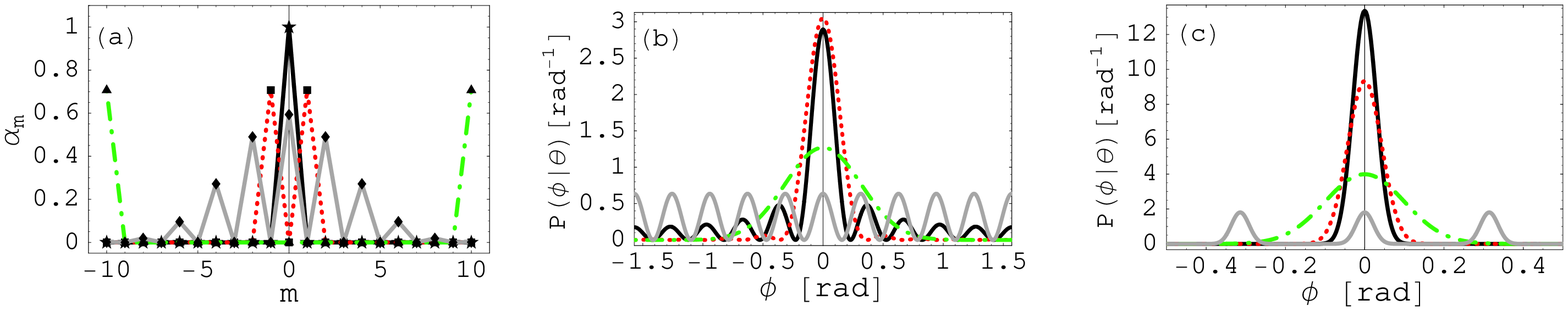}
\caption{(Color online) (a) Probability amplitudes $\alpha_m$ for the twin-Fock
state (solid black line), the external N00N state (green dot-dashed line), the internal N00N state Eq. (31) (solid gray line) and the di-Fock state
$|\psi_{\rm di}(q=1)\rangle$ (red dotted line). Corresponding likelihood
functions for (b) ($M=1)$ and (c) ($M=10)$. In the di-Fock state secondary peaks are
already almost completely absent for $M=1$, thus concentrating
more probability density around $\phi=0$. Secondary features are
however suppressed for larger values of $M$ in all cases,
indicating that the twin-Fock state results in a slightly smaller phase
uncertainty. }\label{knownpsis}
\end{figure*}

\section{Results}

This section summarizes results of a numerical search for optimum
input states of the interferometer. This search employed the
global optimization protocol of simulated annealing
\cite{Bohachevsky1986,Press1999}, whose main features are
summarized in Appendix B.

To set the stage for this discussion, we first recall that several
states have previously been proposed as good candidates for
Heisenberg limited interferometry. One such state is the balanced
twin-Fock input state \cite{Kim1998, Pezze2006}
    \be
    |\psi_{\rm tw}\rangle=|j,0\rangle\equiv
    |N/2\rangle_a|N/2\rangle_b, \label{fock}
    \ee
a state that we use as a benchmark in the following discussion. It
was suggested in Refs. \cite{Yurke1986a, Pezze2006} that
improvements over that state can be achieved by using instead the
state
    \be
    |\psi_{\rm di}(q)\rangle =
    1/\sqrt{2}\left(|j,q\rangle+|j,-q\rangle\right), \label{di}
    \ee
with $q=1$. This state, which we refer to as a di-Fock state in
the following, presents the advantage of suppressing secondary
peaks in the likelihood function, thus concentrating more
probability density around the true value of the phase shift. 

It has also been proposed that Heisenberg limited phase sensitivity
can be achieved with the so-called N00N state \cite{Bollinger1996}
\begin{eqnarray}
\nonumber|\psi_{\rm N00N}\rangle\equiv|\psi_{\rm di}(q=j)\rangle&=& 1/\sqrt{2}\left(|j,j\rangle+|j,-j\rangle\right)\\
\nonumber &=&1/\sqrt{2}\left(|N\rangle_a|0\rangle_b+|0\rangle_a|N\rangle_b\right).\\
\label{noon}
\end{eqnarray}
Some disagreement exists in the current literature regarding the phase sensitivity of N00N states, with some authors claiming that it in fact obeys shot noise limited sensitivity \cite{Pezze2006}.  Mitchell \textit{et al.} \cite{Mitchell2004} as well as Walther \textit{et al.}
\cite{Walther2004} have pointed out that $N00N$ states may be used to produce
super-resolving phase oscillations, with a period of $2\pi/N$, in interferometric
measurements.  In agreement with this we will show that similar
oscillations occur in the likelihood function, \textit{if} Eq. (30)
describes the state of the system \textit{after the first beam
splitter}.  To be explicit we distinguish between external N00N states for which Eq. (\ref{noon}) is the state \textit{before} the first beam splitter, and internal N00N states for which Eq. (\ref{noon}) is the state \textit{after} the first beam splitter.  The internal N00N state is equivalent to using an input state 
\begin{equation}
e^{-i\hat J_x\pi/2}|\psi_{\rm N00N}\rangle.
\label{noonext}
\end{equation}
It can also be achieved by using as an input the state $|N\rangle_a|0\rangle_b$ and replacing the first beam splitter with a non-linear beamsplitter with appropriate interaction time, as shown in \cite{Molmer1999}.

\begin{figure*}
\includegraphics[angle=0, scale=0.6]{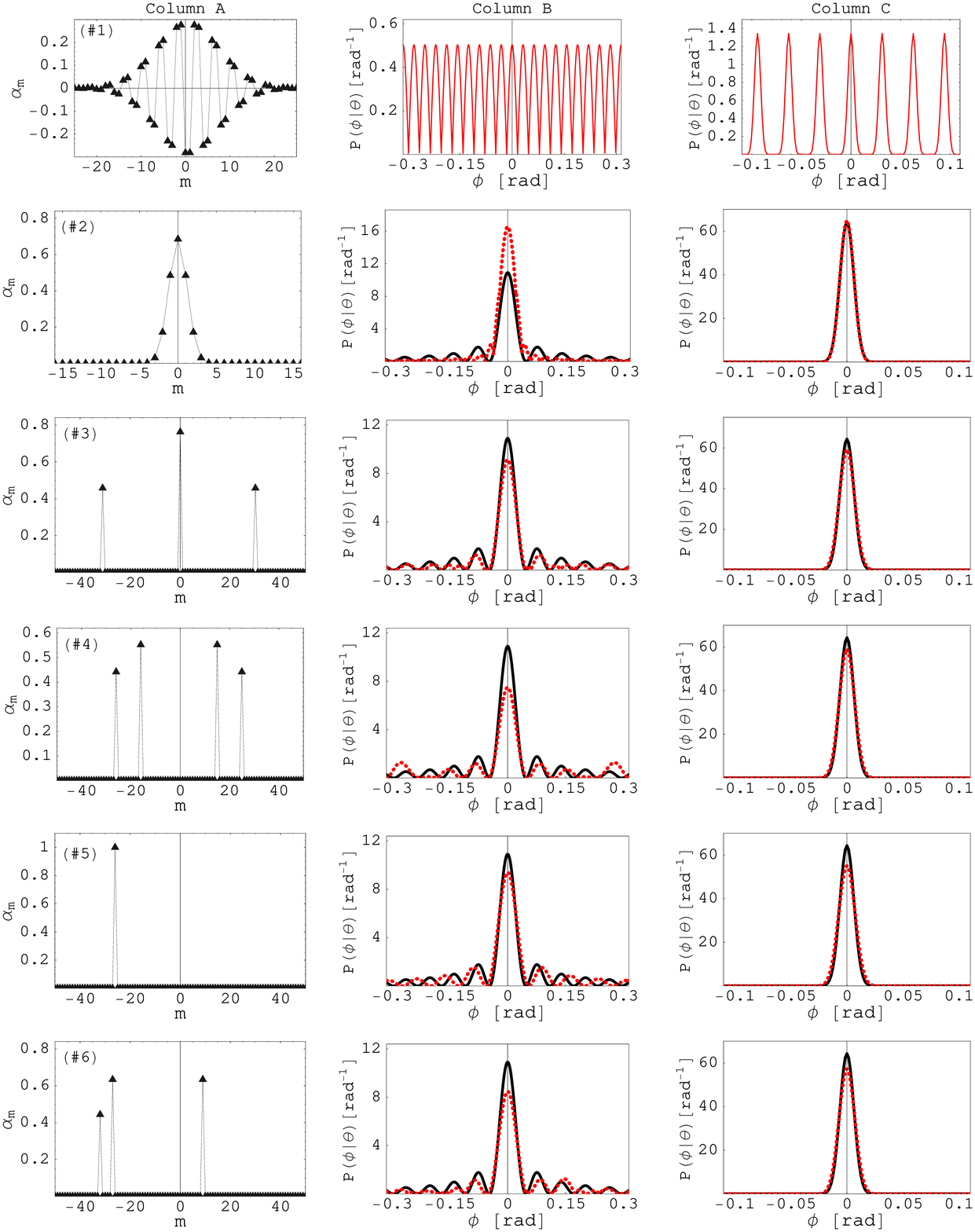}
\caption{(Color online) Column A: probability amplitudes for several possible
input states with $N=100$ particles. Column B: Corresponding
likelihood functions for $M=1$ (dotted red
line), compared to the result for the benchmark twin-Fock input
(solid black line). Column C: Likelihood functions for ($M=10$). }\label{candidates}
\end{figure*}

Figure \ref{knownpsis}(a) shows the probability amplitudes
$\alpha_m$ of the basis states $|j,m\rangle$  for the input states
of Eqs.~(\ref{fock})-(\ref{di}), as well as the external N00N and internal N00N states, for a system with $N=20$ particles, and a
relative phase $\theta=0$ between the two arms of the
interferometer. The solid black line corresponds to the twin-Fock input, the dotted red line to the input state
$|\psi_{\rm di}(q=1)\rangle$, the green dot-dashed line to the external N00N state and the gray solid line to the internal N00N state . The corresponding likelihood functions for $M=1$ are plotted in Fig.
\ref{knownpsis} (b). Apart from the internal N00N state the probability density is
concentrated close to $\phi=0$ in all cases, but the di-Fock state
seems more favorable as it results in a narrow distribution with
no significant secondary peaks. However, this apparent advantage
rapidly disappears  for larger $M$, in which case the secondary peaks associated with the
twin-Fock state are suppressed, leading to a slightly narrower
distribution. Note the distribution corresponding to the external N00N state remains
considerably wider than the other candidates, indicating a
larger uncertainty in the phase estimate.  The likelihood function of the internal N00N state rapidly oscillates with a period of $2\pi/N$ radians.  This is consistent with the $N$-fold increase in phase oscillations observed in \cite{Walther2004,Mitchell2004}.

Figure \ref{candidates} illustrates some of the large number of
possible input states numerically obtained from  the simulated
annealing algorithm for a system of $N=100$ particles. Column A
plots the probability amplitudes $\alpha_m$ of the input states;
column B shows the corresponding likelihood functions with $M=1$
(broken red broken lines) and compares them to the likelihood
function of the benchmark twin-Fock input (solid black lines);
column C plots the situation for $M=10$ measurements. A remarkable
feature of these results is that while these input states are very
markedly different, their likelihood functions become almost
indistinguishable for large $M$. Surprisingly perhaps the
optimization procedure clearly shows the existence of a large
number of local minima resulting in almost identical likelihood
functions.

\begin{figure}
\includegraphics[angle=0, scale=0.4]{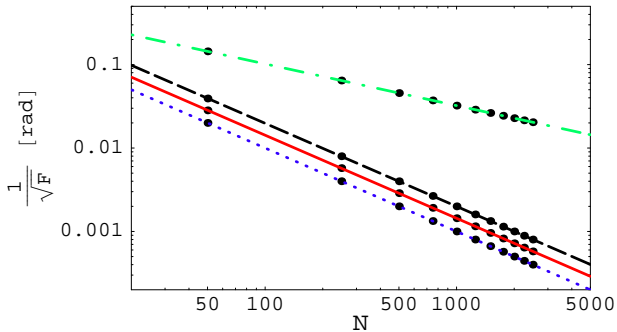}
\label{scaling} \caption{(Color online) Scaling of the inverse square root Fisher
information, $\sigma = 1/\sqrt{F}$ with particle number for the state in row 1 of Fig. 3 (blue dotted line),
twin-Fock state (solid red line), the ``gaussian state" (black dashed line) and the external N00N state (green dot-dashed
line).  The points are calculated using Eq. (\ref{deffisher}) while the lines correspond to least square fits, with the values external N00N state: $\sigma = 1.0/N^{1/2}$, ``gaussian'': $\sigma=1.9/N$, twin-Fock: $\sigma = 1.4/N$ and state in row 1 of Fig. 3: $\sigma=1.0/N$. Note the log-log scale.}
\end{figure}

To demonstrate that all states identified by the simulated
annealing algorithm indeed result in approximate Heisenberg
limited phase sensitivity, Fig. 4 shows a log-log plot of the
inverse square root Fisher information, $1/\sqrt{F}$,  as a
function of particle number over a range of $50\leq N\leq 2500$.
The solid red line is for twin-Fock states, the green dot-dashed
line for external N00N states, the dotted blue line for the uppermost state
in Fig.~\ref{candidates}, and the dashed black line for the state
in row 2 of Fig.~\ref{candidates}. For clarity we refer to the latter
state as a "gaussian state"  in the following. All lines are least
square fits of the equation
    \be
    \sigma = \frac{C}{N^\beta}, \label{HLeq}
    \ee
where $C$ and $\beta$ are fit parameters.

Table 1 summarizes the results of this fit for the states of
Fig.~\ref{candidates}, the number referring to the row number in
the figure. Up to differences of a few percent in the overall
proportionality constant $C$, all of these states clearly satisfy
the $1/N$ scaling characteristic of Heisenberg limited sensitivity,
the only notable exception being the external N00N state, which (in agreement with Pezz\'e and Smerzi \cite{Pezze2006}) is
shot-noise limited.  On the other hand, the inverse square root Fisher information of the
internal $N00N$ state does scale with the Heisenberg limit.  Despite this, the rapid oscillations in the likelihood function seen in Fig. 2(b) and (c) allow a phase estimate only modulo $2\pi/N$.  The consequent ambiguities imply that the internal N00N state may not be useful for phase estimation when using the current Bayesian analysis unless one has \textit{a priori} knowledge that the phase shift lies within a particular phase-bin of width $2\pi/N$.
We now discuss the candidates obtained by search algorithms in turn.

The state with the highest Fisher information that we found, an
apparent global optimum, is shown in row 1 of Fig.
\ref{candidates}. The envelope of its probability amplitudes
$\alpha_i$ is a Gaussian with width $w=\sqrt{N}$.  Despite the
high Fisher information of that state, though, it produces a
significant ambiguity in the determination of the phase estimate
as secondary peaks persist even for $M=10$, as seen in column C. A
feature not apparent on the scale in this figure is that the
central peak is the absolute maximum and becomes increasingly
dominant for increased $M$. Yet, as in the case of the internal N00N state, the persistence of secondary
peaks for relatively large sequences of measurements indicates
that it may not be the most useful state in practice.

In the case of the ``gaussian state", second row of
Fig.~\ref{candidates}, the probability amplitudes $\alpha_m$ have
a Gaussian distribution around the state $|j,m=0\rangle$,
    $$
    \alpha_m = \exp{\left[-m^2/\sigma^{\prime 2}\right]}
    /\cal{N'},
    $$
where $\cal{N}'$ is a normalization constant, and the standard
deviation is $\sigma^\prime=1.7$ for the example at hand. That
state results in Heisenberg limited sensitivity for
$0<\sigma^\prime\lesssim j$, with
    \be \frac{1}{\sqrt{F}} \approx
    \frac{\sigma^\prime}{N}.\label{gausscale}
    \ee
The limit $\sigma^\prime\rightarrow 0$ corresponding to the
twin-Fock state.

\begin{table}
\begin{tabular}{|c|c|}
\hline \textbf{State} & $1/\sqrt{F} = C/N^\beta$ \\ 
\hline\hline $|j,0\rangle$ & $1.4/N$  \\ 
\hline External N00N state & $1.0/N^{1/2}$  \\
\hline Internal N00N state & $1.0/N$  \\
\hline $|\psi_{di}(q=1)\rangle$ & $2.0/N$ \\ 
\hline \#1 &  $1.0/N$\\ 
\hline \#2 & $1.9/N$  \\
\hline \#3 & $1.5/N$  \\
\hline \#4 & $1.5/N$  \\
\hline \#5 & $1.7/N$  \\
 \hline \#6 & $1.6/N$  \\
\hline
\end{tabular}
\caption{Scaling of the phase uncertainty as a function of the
particle number $N$ for the the twin state, the external- and internal N00N states, the
di-Fock state, and the input states of Fig.~\ref{candidates}.}
\end{table}

The state described in the third row of Fig.~\ref{candidates} is
an example of a  state we refer to as a tri-Fock state, and it has the form
    \be \label{trifock}
    |\psi_{\rm tri}(q)\rangle=\left(\alpha_-|j,-q\rangle+\alpha_0|j,0\rangle
    +\alpha_+|j,q\rangle\right)/\cal{N}',
    \ee
where $\cal{N}'$ is a normalization constant.  We find numerically that it results in
Heisenberg limited sensitivity for any value of $q$.

The fourth row in Fig.~\ref{candidates} describes a state that is
a superposition of four Fock states. For the $N=100$ particles
considered in our simulations, and the state in row one of Fig.~\ref{candidates} aside, we have found states with
superpositions containing up to $\sim 8$ Fock states that result
in Heisenberg- limited sensitivity.

We also found that di-Fock states, Eq.~(\ref{di}), with
arbitrary $q$ generally result in
Heisenberg limited or near Heisenberg limited scaling for $q$ as
large as $q\lesssim 0.98 j$. For larger $q$ the state approaches
the shot-noise limited external N00N state. 

Several general trends can be noted in the results of our search.
First, we find that the scaling of Eq.~(\ref{HLeq}) depends only
weakly on the relative probability amplitudes $\alpha_M$ of the
Fock states involved. Changing the relative amplitudes of these
coefficients by factors as large as 3 typically results in changes
in the coefficient $C$ by a few percent only. The Gaussian state
is a notable exception to this trend, and obeys instead the
scaling equation (\ref{gausscale}).

Second, we found no inherent symmetry in the input states that
result in the Heisenberg limit. This is illustrated by the states
of rows 5-6 in Fig.~\ref{candidates}. For example, the state of
row 5 is an unbalanced twin-Fock state of the form
    \be
    |\psi_{\rm tw}\rangle = |j,\gamma j\rangle,
    \ee
where $\gamma$ is some fraction. We found numerically that the state resulted
in Heisenberg limited sensitivity for $0.02\lesssim\gamma\lesssim
0.98$.

All states shown in Fig. \ref{candidates} have real amplitudes.
Allowing for complex amplitudes of the same magnitudes retains
Heisenberg limited or near Heisenberg limited scaling, with
$\beta\gtrsim 0.95$ in Eq. (\ref{HLeq}) and the coefficient $C$
changed by only a few percent. Again the effect is more pronounced
in the Gaussian state, where the change in $C$ can be up to a
factor of $\sim 2$. This is because that state has neighboring
states occupied, and the number statistics of these states
influence each other even for small phase shifts.

Due to the existence of numerous input states resulting in nearly
identical uncertainties within the measurement scheme presented
here, a more relevant criterion for the selection of an
appropriate input state is likely to be its ease of experimental
realization. We address this point in some more detail in the next
section for the case of the "gaussian state."

\begin{figure*}
\includegraphics[angle=0, scale=0.8]{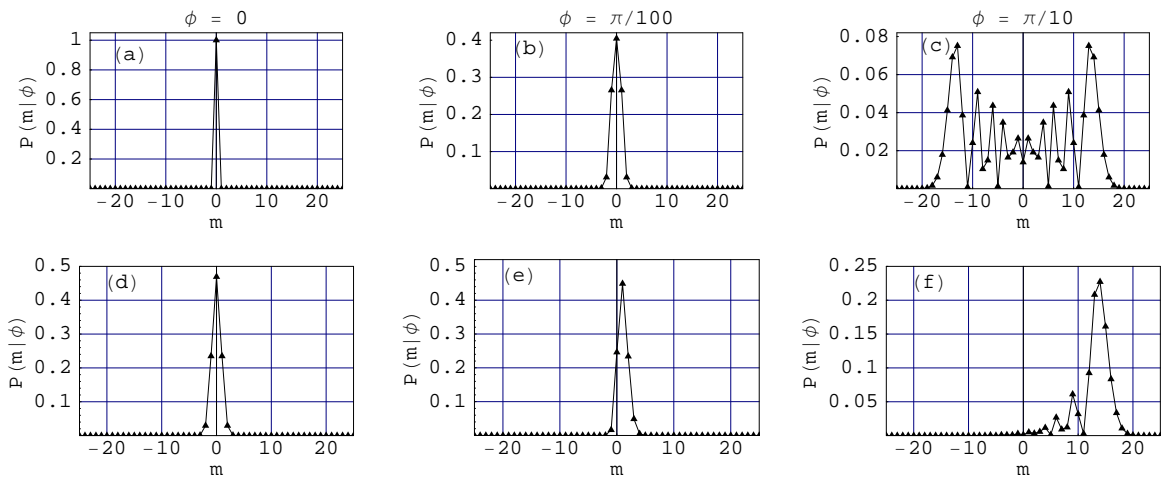}
\label{numstats} \caption{(Color online) Number statistics for phase shifts of
$\phi=0$, $\phi=\pi/100$ and $\phi=\pi/10$ in a system with
$N=100$ particles for (a)-(c) a twin-Fock input state and (d)-(f)
a "gaussian" input state. The twin-Fock state populates number
states symmetrically around $m=0$, leading to an expectation value
of $\langle\hat J_z\rangle=0$ for all values of $\phi$.  This is
not the case for a gaussian state.}
\end{figure*}

\section{The Gaussian state}

The ``gaussian" input state is a promising candidate for
Heisenberg limited interferometry for two reasons: (1) there is a
simple experimental scheme available to generate it;  (2) as we
show below, the expectation value of $\hat J_z$ is phase dependent
in that case (as opposed to the situation for twin-Fock input
states) providing an alternative phase estimate to the direct
measurement of the likelihood function, while still allowing
Heisenberg limited sensitivity.
\subsection{Number statistics}

As mentioned in section III, the likelihood function can be
experimentally reconstructed by multiplying the phase probability
distributions $P(i|\phi)$ associated with a sequence of
measurement outcomes.  This is the approach that was adopted in
the simulations carried out in Refs. \cite{Holland1993,Kim1998}
for twin-Fock input states.  In that case however the average
particle number difference remains zero for all relative phase
shifts between the interferometer arms, and is therefore not a
useful observable. The same is true for the majority of the states
that we identified in our numerical optimization search. One way
to circumvent this difficulty is to measure instead the variance
of $\hat J_z$, an approach that still results in Heisenberg
limited estimates. However, as was pointed out in
Ref.~\cite{Kim1998} for the case of the twin-Fock state, the
signal to noise ratio is then small, $\langle \hat
J_z^2\rangle/\sqrt{\langle \hat J_z^4\rangle -\langle \hat
J_z^2\rangle^2} = \sqrt{2}$.

One advantage of using a gaussian input state instead is that
$\langle \hat J_z \rangle$ now depends on the relative phase
$\phi$. This is illustrated in Fig. 5(a)-(c), which shows the
probability distribution $P(m|\phi)$ at the exit of the
interferometer for twin-Fock and gaussian input states, and for
phase shifts $\phi=0$, $\phi=\pi/100$ and $\phi=\pi/10$. The
expectation value of $\hat J_z$ for the Gaussian state is clearly
not equal to zero for non-zero phase shifts, and may therefore be
used directly to estimate that shift. The uncertainty in $\hat
J_z$, evaluated via the standard error propagation formula
Eq. (\ref{standerr}), is shown in Fig.
6 as a function of the number of interfering particles.  Least
square fits indicate that the uncertainties are Heisenberg
limited, with $\Delta\phi = 2.0/N$ for the gaussian state  and
similarly $\Delta\phi = 2.4/N$ in the case of a tri-Fock state
with $q=1$.

\begin{figure}
\includegraphics[angle=0, scale=0.38]{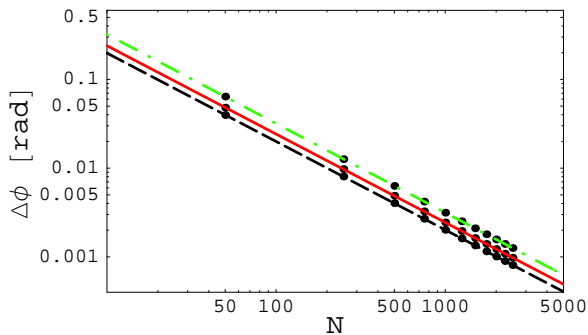}
\label{scalenormerr} \caption{(Color online) Scaling of the uncertainty in the
estimate of the relative phase from a measurement of average
particle number difference.  The dashed black line is for a
Gaussian state and the red solid line of a tri-Fock state, Eq.
(\ref{trifock}), with $q=1$, and the green dot-dashed line
represents the state engineered as described in section V(B). The
points are numerically determined while the lines correspond to
least square fits. Note the log-log scale.}
\end{figure}

\subsection{Input state engineering}

We have seen that the constraints on the relative phases of the
complex probability amplitudes $\alpha_i$ of the input states are
surprisingly weak when estimating phase shifts $\phi$ via a
reconstruction of the likelihood function. However, the number
statistics of the field after passage through the interferometer
depend critically on these relative phases. For example, if in the
Gaussian state of Fig. \ref{candidates} the components
$|j,-1\rangle$ and $|j,+1\rangle$ are $\pi/2$ radians out of phase
with $|j,0\rangle$, the output state $|\psi_{\rm out}\rangle$ will
be populated symmetrically around $|j,0\rangle$ independently of
$\phi$, so that $\langle\hat J_z\rangle=0$ for all $\phi$. Hence,
some care must be taken in preparing the input states $|\psi_{\rm
in}\rangle$.

It is possible to generate a gaussian input state from an initial
twin-Fock state, Eq. (\ref{fock}), by subjecting it to the
Hamiltonian
    \be
    H_x = \hbar g\hat J_x,
    \ee
where $g$ is a coupling constant, for a time $\tau = 3\pi/4Ng$.
The resulting state is precisely the state shown in row 2 of Fig.
3, but with the probability amplitudes of the components
$|j,-1\rangle$ and $|j,+1\rangle$ $\pi/2$ radians out of phase
with the component $|j,0\rangle$. These three components can be
brought into phase via an additional evolution under the
Hamiltonian
    \be \label{nonlinH}
    H_z = \hbar \chi\hat J_z^2
    \ee
for a time $\tau = 3\pi/2\chi$. The resulting state allows Heisenberg limited phase estimation by measuring
$\langle\hat J_z\rangle$, with the scaling law $\Delta\phi =
3.2/N$ as a function of the particle number $N$. We remark that
while the time evolution (\ref{nonlinH}) of the input state brings
the three main components of $|\psi_{\rm in}\rangle$ in phase with each
other, that is not so for the other, weakly populated number
states that comprise it. This results in a somewhat reduced
performance compared to Gaussian state with all components in
phase.

Hamiltonians of the form Eq. (\ref{nonlinH}) have long been known
to act as squeezing operators in interferometers
\cite{Kitagawa1993}.  In the case of photons, they can be
implemented by inserting an optical Kerr medium into each arm of
the interferometer \cite{Kitagawa1986}. In the case of charged particles they arise
due to mutual phase modulation from Coulomb interaction between
particles in each arm \cite{Kitagawa1991}. Atomic spins coupled to
the polarization of an optical field \cite{Smith2006,Geremia2006}
can lead to similar Hamiltonians for neutral atoms.

\section{Conclusion}

We have used a global optimization scheme to systematically search
for input states of a quantum mechanical Mach-Zehnder
interferometer that yield phase estimates with accuracy scaling
like the inverse number of particles, the Heisenberg limit.
Surprisingly perhaps, we find that a large number of states can
achieve that limit. They typically consist of superpositions of a
small number of Fock states, with few restrictions on the relative
phase of their complex amplitudes on on their symmetry. An input
state of particular relevance consists of a gaussian distribution
of amplitudes around the state $|N/2\rangle_a|N/2\rangle_b$, due
principally to its relative ease of realization with simple
Hamiltonian dynamics.

\acknowledgements
 We would like to thank Lajos Diosi for
insightful conversations regarding the relation between the
likelihood function and information theoretical concepts. This
work is supported in part by the US Office of Naval Research, by
the National Science Foundation, by the US Army Research Office,
and by the National Aeronautics and Space Administration.

\appendix

\section{Fisher information}


Consider the deviation $\langle \delta A_\theta\rangle$ of an
observable from its mean value $\langle A(\theta)\rangle$  at a fixed phase $\theta$, as a
function of the phase shift $\phi$,
    \be
    \langle \delta A_\theta\rangle = \sum\limits_j \left(a_j-\langle
    A(\theta)\rangle\right)P(j|\phi).
    \ee
Then
\begin{eqnarray}
\frac{d\langle \delta A_\theta\rangle}{d\phi} &=&
\sum\limits_j \left(a_j-\langle A(\theta)\rangle\right)
\frac{dP(j|\phi)}{d\phi}\label{difdelA}\\
&=& \sum\limits_j \left(a_j-\langle A(\theta)\rangle\right)
P(j|\phi)\frac{\log{P(j|\phi)}}{d\phi}\label{DA}.
\end{eqnarray}
Regrouping the factors in Eq. (\ref{DA}) and squaring gives
\begin{widetext}
\begin{eqnarray}
\left[\frac{d\langle \delta A_\theta\rangle}{d\phi}\right]^2&=&
\left[\sum\limits_j \left(\left(a_j-\langle A(\theta)\rangle\right)
\sqrt{P(j|\phi)}\right)\left(\sqrt{P(j|\phi)}\frac{\log{P(j|\phi)}}
{d\phi}\right)\right]^2\\
&\leq& \left[\sum\limits_j P(j|\phi)\left(\frac{\log{P(j|\phi)}}
{d\phi}\right)^2\right]\left[ \sum\limits_j \left(a_j- \langle
A(\theta)\rangle\right)^2P(j|\phi)\right],
\label{schwarz}
\end{eqnarray}
\end{widetext}
where we have used the Schwarz inequality in the last step. Noting
that the second term in Eq. (\ref{difdelA}) vanishes due to
normalization we have $d\langle \delta
A_\theta\rangle/d\phi=d\langle A\rangle/d\phi$. We can then
rewrite inequality (\ref{schwarz}) and evaluate it at $\theta$ to
give the desired result: \be \Delta\phi^2=\frac{\Delta
A^2}{\left[d\langle A\rangle/d\phi\right]^2}\geq
\frac{1}{\sum\limits_j
P(j|\theta)\left(\frac{\log{P(j|\phi)}}{d\phi}\arrowvert_\theta\right)^2}.
\label{crmroaA} \ee This is the Cramer-Roa inequality, Eq.
(\ref{cramerrao}), in the current context.  The denominator on the
right in Eq. (\ref{crmroaA}) defines the Fisher information.

\section{Simulated annealing}

Simulated annealing \cite{Bohachevsky1986, Press1999} is a
mathematical approach to global optimization simulating the
metallurgical process whereby an amorphous compound is
successively heated and cooled while gradually lowering the
average temperature in an attempt
to enlarge the grain size of single crystals in the compound.  The protocol is as follows:\\
1. Initial conditions for the
optimization parameters, $\mb{x}=(x_1,x_2,...x_r)$ are chosen, usually at random.\\
2. With each allowed value of $\mb{x}$ is associated an pseudo-energy,
$E(\mb{x})$, which
is the quantity to be minimized.\\
3. A new value, $\mb{x}^\prime=\mb{x}+\mb{\Delta x}$, is then
generated for the optimization parameters,
and the change in energy $\Delta E=E(\mb{x}^\prime)-E(\mb{x})$ calculated.\\
4. The new value of $\mb{x}$ always replaces the old if
$p=e^{-\Delta E/kT} > 1$, or with probability $p$ if $p<1$.
Here $k$ is a constant of proportionality  analogous to Boltzmann's constant
in statistical mechanics, and $T$ is a pseudo-temperature.\\
5. The process is repeated.\\

Accepting new parameter sets for which $p<1$, i.e. uphill steps on
the energy manifold, allows the algorithm to explore the whole
parameter space instead of converging directly to the closest
local minimum.

Two important considerations in this procedure are the method of
choosing the next set of parameters $\mb{x}^\prime$ and the
annealing schedule, i.e. the protocol for gradually lowering the
temperature with intermittent heating cycles until the system has
frozen into, hopefully, a global minimum. Various approaches to
these considerations have been discussed in the literature
\cite{Vanderbilt1984,Bohachevsky1986}.

In our implementation the optimization parameters are the set of
amplitudes $\alpha_m$ of the input state vector Eq.
(\ref{instate}) and the energy the inverse square root of the
Fisher information.  We execute the simulated annealing algorithm
not on a single vector $\mb{\bar \alpha}$, but a population of
vectors chosen at random.  The pseudo-temperature of the system is
set by the average energy of the population,
    \be
    T=\frac{1}{P}\sum\limits_{i=1}^P 1/\sqrt{F_i},
    \ee
where $P$ is the number of state vectors in the population and
$F_i$ the Fisher information of the $i^{\rm th}$ state vector.
Defining the temperature in this way self-regulates the cooling
cycle. If a single global minimum exists, the algorithm will
continue sampling until the majority of state vectors have fallen
into the global minimum.  On the other hand if many local minima
of comparable depth exist the algorithm will also continue to
sample the parameter space until the majority of state vectors
have found such local minima. As more local minima are found the
system "cools down" by itself.

When the state vectors $\bar\alpha$ have converged near the minima
and the step size
$|\Delta\bar\alpha|=|\bar\alpha^\prime-\bar\alpha|$ is fixed, all
new steps $\bar\alpha^\prime$ will be uphill, thus halting further
convergence.  To enable further convergence we therefore half the
step size when the number of downhill steps found over a several
iterations drops below a threshold.

It may also happen in our approach that the system reaches an
equilibrium condition in which the average number of uphill steps
accepted become equal to the average number of downhill steps
found.  To ensure that the system continues to converge towards
minima, we lower Boltzmann's constant by $k\rightarrow 0.9 k$ if
this point is reached.  We take as an indicator that the system is
near this point whenever the number of accepted uphill steps is
larger than the number of downhill steps in a given iteration
cycle.

To summarize the algorithm:
\begin{enumerate}
\item Choose initial population at random and calculate
pseudo-temperature \item Find new state vectors
$\bar\alpha^\prime=\bar\alpha+\Delta\bar\alpha$ and replaces the
old if $p=e^{-\Delta E/kT} > 1$, or with probability $p$ if $p<1$
\item If the number of uphill steps accepted is greater than
number of downhill steps decrease $k\rightarrow 0.9 k$. \item If
the number of downhill steps found is less than specified
threshold reduce $\Delta\bar\alpha\rightarrow 0.5
\Delta\bar\alpha$. \item Repeat algorithm
\end{enumerate}

We have implemented searches that assume either real or complex
amplitudes. In addition, in some searches we imposed no
restrictions on the symmetry of input states, while in others we
forced the input states to be either symmetric or anti-symmetric
around $|N/2\rangle_a|N/2\rangle_b$.

In the case of real amplitudes, the initial population was chosen
by generating a random number between $[-1,1]$ for each $\alpha_m$
and then normalizing the state vector.  For complex amplitudes the
magnitudes $|\alpha_m|$ were chosen at random between 0 and 1
and a complex phase between 0 and 2$\pi$.

We have used two different approaches to specifying the new state
vectors $\bar{\alpha}^\prime$ during each iteration. In the first
one new state vectors were selected by changing each amplitude at
random within an interval
$\epsilon\alpha_m<\alpha^\prime_m<(1+\epsilon)\alpha_m$, with
$\epsilon<1$ and then renormalizing the state vector.  In the case
of complex amplitudes a new phase was also chosen as
$\phi_m^\prime=\phi_m+\Delta\phi$ where
$-\varepsilon<\Delta\phi<\varepsilon$, while in the case of real
amplitudes sign changes were allowed if $|\alpha_m|<0.02$ by
choosing
$-(1+\epsilon)|\alpha_m|<\alpha^\prime_m<(1+\epsilon)|\alpha_m|$.
In this approach, step (4) in the algorithm described above was
rather insensitive to the values chosen for $\epsilon$ and
$\varepsilon$.  They were therefore taken to have fixed values
$\epsilon=0.5$ and $\varepsilon=0.05$.

In the second approach each vector $\mb{\bar \alpha}$ is specified
by a set of angles such that the vector moves on a hyperspherical
surface of radius $1$ to ensure normalization.  The next vector is
chosen in a random direction on the hyper-sphere with the initial
step size $\Delta\bar\theta=0.1$.  It is decreased in successive
iterations according to step (5) in the algorithm above.

In the first approach the state vectors in the population settle
in many local minima of comparable depths, while in the second
approach all state vectors converge to an apparent global minimum
which is the state shown in row 1 of column A in Fig. 3.


\end{document}